\documentclass[epj]{svjour}
\usepackage{graphicx}
\usepackage{dcolumn}
\usepackage{amsmath}
\usepackage{array}
\usepackage{bm}
\usepackage{amssymb}
\usepackage{amsfonts}

\begin{document}
\title{Glueballs, gluon condensate, and pure glue QCD below $T_c$}

\author{Fabien Buisseret\inst{1}
\thanks{F.R.S.-FNRS Research Fellow; fabien.buisseret@umons.ac.be}%
}                     
\offprints{}          
\institute{Service de Physique Nucl\'{e}aire et Subnucl\'eaire,
Universit\'{e} de Mons -- UMONS, Place du Parc 20, 7000 Mons, Belgium}
\date{Received: date / Revised version: date}
%
\abstract{
A quasiparticle description of pure glue QCD thermodynamics at $T\lesssim T_c$ is proposed and compared to recent lattice data. Given that a gas of glueballs with constant mass cannot quantitatively reproduce the early stages of the deconfinement phase transition, the problem is to identify a relevant mechanism leading to the observed sudden increase of the pressure, trace anomaly, \textit{etc.}~near the critical temperature. It is shown that the strong decrease of the gluon condensate at $T\lesssim T_c$ combined with the increasing thermal width of the lightest glueballs might be the trigger of the phase transition. 
} 

\maketitle

\section{Introduction}
The phenomenology related to the deconfinement phase transition from hadronic matter to quark-gluon plasma (QGP) at high enough temperatures or densities is a very active field of research since the beginnings of QCD, with pioneering works such as Refs.~\cite{coll75,shur80}. Much effort is devoted to understand the QGP on the experimental side: The QCD matter is or will be studied in heavy-ion collisions at RHIC, SPS, FAIR, and the LHC (see Refs.~\cite{ente,fair} for more information). On the theoretical side, the QGP is also a challenging topic, which has deserved lots of studies within several frameworks. In particular, phenomenological quasiparticle models and more fundamental approaches like perturbative calculations or AdS/CFT duality have proved to be successful both in reproducing lattice data and in providing a reliable description of experimentally observed results at RHIC, see \textit{e.g.} Refs.~\cite{peshier,pertu,panero} for some recent results and Refs.~\cite{rev1} for some reviews on the topic. 

Lattice QCD has a particular status since it is, in principle, the most powerful technique to extract nonperturbative informations from QCD. When experimental data are lacking, lattice data are then often used to fit other model's parameters. A crucial observable in QGP physics that can be computed in lattice QCD is the QCD equation of state (EoS). In particular, the EoS of pure glue SU(3) QCD has been computed on the lattice more than a decade ago~\cite{boyd95}, while more detailed data have recently been obtained in Ref.~\cite{panero} with gauge groups ranging from SU(3) to SU(8). Many other calculations of the QCD EoS have also been performed including quarks flavors, at zero chemical potential or not. The interested reader will find more references in \textit{e.g.} the review~\cite{rev2}. 

Among the various existing phenomenological approaches, quasiparticle models rely on the assumption that the QGP can be seen as a gas of deconfined quarks and gluons. Since pioneering works~\cite{golo}, they have been shown to be able to reproduce the various EoS computed in lattice QCD at $T\geq T_c$. It is thus tempting to apply such models below $T_c$. In that case however, the QCD matter should rather be modeled by a hadron gas since QCD it is in its confining phase. We focus here on the pure glue EoS below $T_c$, which already captures the essential physical features of the full QGP without involving extra technical difficulties due to the presence of quarks. 

\section{Glueball gas model}\label{ggm}
The pure glue hadronic matter is expected to consist in glueballs and, since the scattering amplitudes between glueballs scales in $1/N_c^2$ (instead of $1/N_c$ for mesons), a noninteracting Bose gas may be assumed in a first approach. Following Bose-Einstein statistics, only the lowest-lying glueballs should bring a significant contribution to the EoS since the one of a glueball of mass $m_g$ and spin $J_g$ is roughly proportional to $(2J_g+1){\rm e}^{-m_g/T}$. However, by using typical values for the low-lying glueball masses~\cite{glus} in a hadron gas model, one fails to reproduce the strong increase of the thermodynamical variables near $T_c$, as already pointed out in Refs.~\cite{suga1,meyer}. To our knowledge, the only proposal leading to a model in agreement with the lattice EoS is the one of Ref.~\cite{meyer}, where a high-lying glueball spectrum of Hagedorn-type is assumed. The exponential growth of the number of states with mass $m_g$ then compensates the statistical suppression and brings a large enough contribution near $T_c$. It is also worth mentioning the recent Ref.~\cite{hage}, where it is shown that a Hagedorn-type model of the QGP is able to reproduce the trace anomaly computed in lattice QCD while ensuring a low viscosity/entropy ratio.  

The Hagedorn spectrum relies on a string-theoretical picture of the hadron spectrum. In particular, following the ideas of Ref.~\cite{meyer}, one has to assume that glueballs are excitations of a closed string, which is not so obvious, partly in view of the many models reproducing the lattice spectrum at zero temperature by assuming totally different frameworks~\cite{revvin}. That is why it might be of interest to see whether an alternative way of understanding the early stages of the pure glue phase transition ($T\lesssim T_c$) can be found or not. This is the purpose of the present work, where it is chosen not to assume a Hagedorn glueball spectrum. Therefore, only the lightest glueballs, that all models find to be the scalar and tensor ones in pure glue QCD~\cite{revvin}, are assumed to bring a significant contribution to the EoS. 

As it will be shown, two ingredients are needed. First, the nontrivial contribution of the gluon condensate to the trace anomaly, especially its brutal reduction near the phase transition. Remark that the strong link between the gluon condensate and the QCD phase transition have been previously outlined for example in Refs.~\cite{simonov}. Second, a significant reduction of the glueball masses near $T_c$. Since that point is the most counterintuitive one at first sight, let us analyze it first.  

\section{Glueball mass reduction}\label{mred}

The behavior of glueballs at finite temperature has been first investigated on the lattice in Refs.~\cite{suga2,suga3}. The basic conclusion of those works is twofold, following the procedure used to analyze the results. On one hand, if the temporal glueball correlator is fitted assuming glueball states with zero width, then the $0^{++}$ and $2^{++}$ pole masses are found to significantly decrease for $T\lesssim T_c$. On the other hand, a Breit-Wigner fit can be applied to that correlator; then the glueball masses are found to be roughly constant from $T=0$ to $T_c$, with a thermal width which is nearly zero at low temperature but then increases rather linearly near $T_c$. That conclusion is quite intuitive and might be a general feature of glueballs near $T_c$: The glueballs should progressively be ``dissolved" in the medium when approaching the deconfinement temperature, and their width should increase accordingly. Notice that such a constant mass is expected from calculations relying on effective low-energy Lagrangians~\cite{lowe}. It is moreover shown in Ref.~\cite{suga3} that the pole mass, $m_g(T)$, and the Breit-Wigner mass, $\bar m_g(T)$, and thermal width, $\Gamma_g(T)$, are linked as follows:
\begin{subequations}\label{mfit}
\begin{equation}\label{glum}
	m_g(T)\approx \bar m_g(T)-2T+\sqrt{4T^2-\Gamma_g(T)^2}.
\end{equation}
\begin{figure}[t]
\includegraphics*[width=8.0cm]{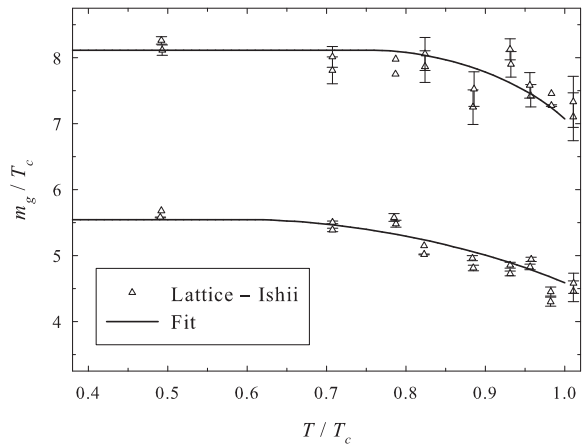}
\caption{Scalar and tensor glueball masses computed through a pole-mass fit in Ref.~\cite{suga3} (symbols), compared to a fit of the form~(\ref{mfit}) with (in units of $T_c$) $T_{0^{++}}=0.596$, $m^0_{0^{++}}=5.547$, $b_{0^{++}}=4.230$, $T_{2^{++}}=0.755$, $m^0_{2^{++}}=8.113$, and $b_{2^{++}}=7.152$. A standard value $T_c=265$~MeV is used.}
\label{fig1}
\end{figure}
As shown in Fig.~\ref{fig1}, the pole masses computed in Ref.~\cite{suga3} are well described below $T_c$ by the form (\ref{glum}) with
\begin{align}
	\bar m_g(T)&=m^0_g, \nonumber\\
	\Gamma_g(T)&=0  & T\leq T_g\nonumber\\
	&=b_g\, (T-T_g)& T_g<T<T_c.
\end{align}
\end{subequations}
Such an ansatz qualitatively encodes the results of the Breit-Wigner fit performed in Ref.~\cite{suga3}
. Remark that the thermal broadening of the glueballs generates a pole-mass reduction; interestingly such a glueball mass reduction near the phase transition has also been predicted by computations within the dual Ginzburg-Landau theory~\cite{suga4}. We point out that the value $T_c=265$~MeV will be assumed in the rest of this paper; it lies in the typical range $260-280$~MeV found in pure glue QCD.  

\section{Glueball gas pressure}\label{glupres}

As said in Sec.~\ref{ggm}, we assume the interactions between glueballs to be negligible in a first approximation. The pressure of an noninteracting gas of glueballs with mass $m_g$ and spin $J_g$ simply corresponds to that of an ideal Bose-Einstein gas~\cite{wal} 
\begin{equation}\label{press}
	p_g=-\frac{(2J_g+1)T}{2\pi^2}\int^\infty_0dk\, k^2 \ln\left(1-{\rm e}^{-\sqrt{k^2+m_g^2}/T}\right).
\end{equation}

Such a framework is the most straightforward way to deal with a boson gas. However, Eq.~(\ref{press}) shows that only the glueball masses appear in the computation of the EoS, and we have seen in the previous section that the increasing thermal width should also be taken into account. Rigorously, one would thus need a statistical formalism not only generalizing the Bose-Einstein distribution to the case of unstable particles, as done first in~\cite{weldon}, but also explicitly including the interactions between glueballs, since an increasing width should enhance those interactions. Such a formalism is out of the scope of the present paper, which should be seen as a first step towards an alternative way of understanding the pure glue EoS below $T_c$. That is why two assumptions will be made in the following. First, we still choose to neglect the interactions between glueballs. As discussed in the introduction, large-$N_c$ arguments might be relevant to justify that point. Second, we consider that the nonzero-width effects can be absorbed into a redefinition of the glueball masses. Indeed, we have shown in the previous section that a glueball with a constant mass and an increasing thermal decay width can be described effectively as a glueball with a zero width and a decreasing pole mass. That is why the pole mass $m_g(T)$ will be used hereafter. Notice that the second approximation is expected to be less and less accurate when approaching $T_c$ from below, since the ratio $\Gamma_g(T)/\bar m_g(T)$ is an increasing function of $T$. From~\cite{suga3} one can compute that $\Gamma_{0^{++}}(T_c)/\bar m_{0^{++}}\approx 0.34$ and $\Gamma_{2^{++}}(T_c)/\bar m_{2^{++}}\approx 0.25$, meaning that the glueball decay widths become significant with respect to their masses at the phase transition. Nevertheless one can reasonably think that our approach will be able to predict the global trend of the EoS near $T_c$. 

Starting from Eqs.~(\ref{press}) in which $m_g=m_g(T)$ given by~(\ref{glum}), the pressure of the QCD matter below $T_c$ should then be given by $ p=\sum_g p_g$, the sum running over all the glueball states. But, since a Hagedorn spectrum is not assumed, only the lowest-lying glueballs will significantly contribute because the statistical suppression in $(2J_g+1){\rm e}^{-m_g/T}$ is not balanced by the exponentially rising number of states with respect to $m_g$. Moreover, the higher mass glueballs would contribute if $m_g/T$ was of order unity, but $m_g/T_c\gg 1$ even for the $0^{++}$ glueball. We thus take 
\begin{equation}\label{press2}
	p\approx p_{0^{++}}+p_{2^{++}}.
\end{equation}

\begin{figure}[t]
\includegraphics*[width=8.0cm]{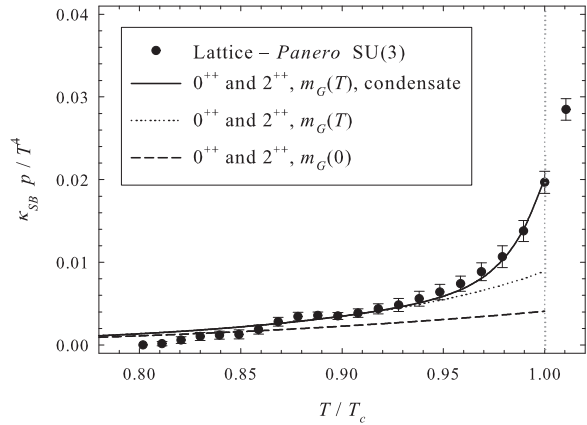}
\caption{Pressure computed in pure glue SU(3) lattice QCD and normalized to the Stefan-Boltzmann factor $\kappa_{SB}=45/8\pi^2$, taken from Ref.~\cite{panero} (full circles). The lattice data are compared to Eq.~(\ref{press2}) using either $m_g=m_g^0$ (dashed line), $m_g=m_g(T)$ given by Eq.~(\ref{mfit}) (dotted line), and the full model in which the gluon condensate contribution (\ref{gluconpress}) is added (solid line).}
\label{fig2}
\end{figure}

Results obtained from this last equation are shown in Fig.~\ref{fig2} and compared to the lattice data of Ref.~\cite{panero}. The conclusions are the following. First, using a glueball gas with constant masses $m_g=m^0_g$ fails to reproduce the observed increase of pressure near $T_c$, even by taking lighter values than the fitted ones, given in Fig.~\ref{fig1}. Second, using the temperature-dependent masses, $m_g=m_g(T)$, fitted from Ref.~\cite{suga3}, greatly improves the agreement with lattice QCD and is a first argument in favor of the scenario proposed here. Remark that $p(T_c)$ is underestimated; we will show in the next section that the contribution coming from the gluon condensate is able to cure this problem. Another possibility could be that the higher-lying glueballs bring a significant contribution to the pressure. Provided that the mass reduction mechanism does not cause those glueballs to become lighter than the tensor one, it can be computed that the $0^{-+}$ contribution shifts the pressure of less than $7\%$ (the $0^{-+}$ state is the closest in mass of the $2^{++}$), while the heavier states contribution is even smaller. Thus only the scalar and tensor glueballs may be considered in a first approximation. It is worth mentioning the recent work~\cite{meng} suggesting that, indeed, the $0^{-+}$ glueball remains heavier than the tensor one. 

\section{Gluon condensate contribution}\label{gcta}
\begin{figure}[t]
\includegraphics*[width=8.0cm]{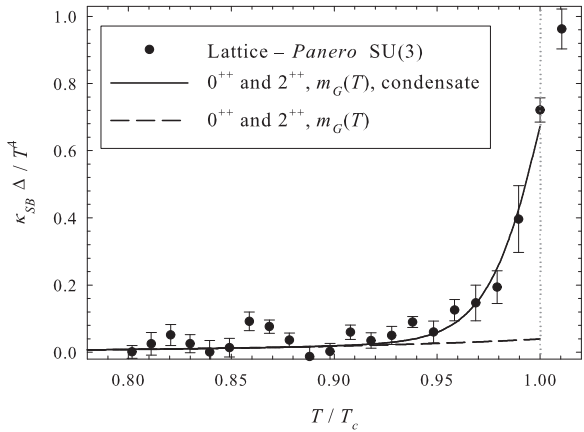}
\caption{Same as Fig.~\ref{fig2} for the trace anomaly. The dashed line is the glueball contribution~(\ref{ta}), while the solid line comes from Eq.~(\ref{ta2}), where $\Delta_{G^2}=\tilde\Delta_{G^2}$ following Eq.~(\ref{ta4}).}
\label{fig3}
\end{figure}

The next step is now the computation of the trace anomaly, \textit{a priori} straightforwardly defined from the pressure~(\ref{press2}) by
\begin{equation}\label{ta}
	\bar\Delta=T^5\partial_T\left(\frac{p}{T^4}\right).
\end{equation}
A look at Fig.~\ref{fig3} clearly shows that our model with $m_g(T)$, although satisfactorily reproducing the lattice pressure, severely underestimates the trace anomaly near the phase transition. This seemingly paradoxical situation can be clarified as follows: When speaking of the trace anomaly, the nontrivial role of the gluon condensate has to be taken into account. That argument has already been proposed when studying the QCD matter at $T>1.2\, T_c$~\cite{castorina}. Let us now apply it below $T_c$. 

It is known that the gluon condensate at temperature $T$, defined by $\left\langle G^2\right\rangle_T=-\left\langle \frac{\beta}{g}G^a_{\mu\nu}G^{\mu\nu}_a(T)\right\rangle$, contributes to the QCD trace anomaly as $\Delta_{G^2}=\left\langle G^2\right\rangle_0-\left\langle G^2\right\rangle_T$~\cite{castorina,leut}. Thus the total trace anomaly, $\Delta$, should rather be
\begin{equation}\label{ta2}
	\Delta=\bar\Delta+\Delta_{G^2}.
\end{equation}
The gluon condensate can moreover be written as the sum of a magnetic and an electric part, \textit{i.e.} $\left\langle G^2\right\rangle_T=\left\langle G^2_e\right\rangle_T+\left\langle G^2_m \right\rangle_T$ in euclidean space. It appears from lattice QCD simulations that $\left\langle G^2_m\right\rangle_T\approx\left\langle G^2_m\right\rangle_0$ on one hand, and that $\left\langle G^2_e\right\rangle_T$ is such $\left\langle G^2_e\right\rangle_0\approx\left\langle G^2\right\rangle_0/2$ but then falls very quickly near $T_c$ to reach a zero value just after the phase transition~\cite{delia}. Consequently, one expects~\cite{castorina} 
\begin{eqnarray}\label{ta3}
	\Delta_{G^2}&=&\frac{\left\langle G^2\right\rangle_0}{2}\left[1-c_e(T)\right],\ {\rm where}\  c_e(T)=\frac{\left\langle G^2_e\right\rangle_T}{\left\langle G^2_e\right\rangle_0}
\end{eqnarray}
can be known from lattice computations~\cite{delia}.

\begin{figure}[t]
\includegraphics*[width=8.0cm]{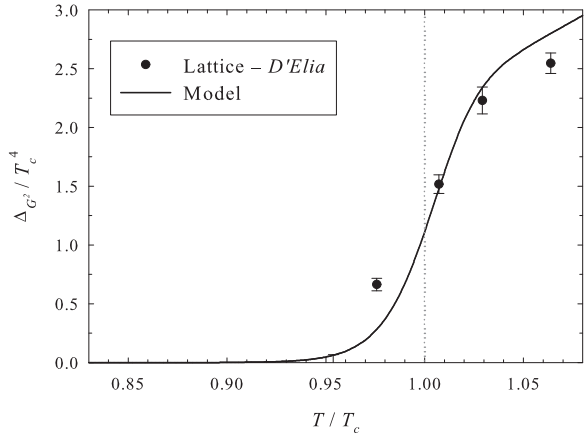}
\caption{Trace anomaly contribution $\tilde\Delta_{G^2}$ needed to fit the lattice data presented in Fig.~\ref{fig3} (solid line). The curve agrees with formula~(\ref{ta3}) in which $c_e(T)$ has been taken from the lattice study~\cite{delia}, together with the typical value $\left\langle G^2\right\rangle_0=$0.030~GeV$^4=$6.08$\, T_c^4$ (full circles).}
\label{fig4}
\end{figure}

Since $c_e(T)$ is only known at a few temperatures from the lattice (see Fig.~\ref{fig4}), it would not be convenient to use the available data as an input in our model. Rather, it is better to compute first the values of $\Delta_{G^2}$ that fit the trace anomaly obtained in Ref.~\cite{panero} and then to check whether they agree or not with the results of Ref.~\cite{delia}. The values needed to fit the trace anomaly computed in Ref.~\cite{panero} is very accurately fitted by the form
\begin{subequations}\label{ta4}
\begin{eqnarray}
	\tilde\Delta_{G^2}&=&f_1\, T^4_c\frac{T}{T_c}\left[\frac{T}{T_c}-f_2\right]\times\nonumber\\&& \left[1-\frac{1}{\left(1+{\rm e}^{(T/T_c-f_3)/f_4}\right)^2}\right]\ \ T\geq 0.9\, T_c,\nonumber\\
	&=&0\hspace{4.2cm} T<0.9\, T_c,
\end{eqnarray}
with
\begin{equation}	
f_1=6.171,\, f_2=0.637,\, f_3=1.013,\, f_4=0.015.
\end{equation}
\end{subequations}

As shown in Fig.~\ref{fig3}, the trace anomaly~(\ref{ta2}) computed with $\Delta_{G^2}=\tilde\Delta_{G^2}$ fits the lattice data very well -- it is logical since it has been designed for. The key observation is now that, in Fig.~\ref{fig4}, the fitted term~(\ref{ta4}) is in good agreement with that obtained by putting the values of $c_e(T)$ available from the lattice study~\cite{delia} in the theoretical estimate~(\ref{ta3}). A typical value $\left\langle G^2\right\rangle_0=$0.030~GeV$^4$ has been used as a mean value of various results obtained so far in the literature~\cite{rakow}. Such an agreement between the needed value of the gluon condensate and the one theoretically expected is a relevant check of the mechanism presented here describing the phase transition.  

The coherence of our scenario requires the gluon condensate contribution to the pressure to be computed from the thermodynamical relation~(\ref{ta}) and then to be added to the glueballs pressure $p$. One has
\begin{equation}\label{gluconpress}
	p_{G^2}=T^4\int^T_{0}\frac{\tilde\Delta_{G^2}(x)}{x^5}\, dx,
\end{equation}
and the total pressure finally reads 
\begin{equation}\label{ptot}
	p=p_{0^{++}}+p_{2^{++}}+p_{G^2}.
\end{equation}
It is readily observed in Fig.~\ref{fig2} that, near $T_c$, the total pressure~(\ref{ptot}) reaches an excellent agreement with the lattice data. In particular, the pressure in $T=T_c$ is no longer underestimated. 
  
Since in Ref.~\cite{panero}, to which our model is compared, the energy density as well as the entropy density are obtained as linear combinations of the pressure and trace anomaly following standard thermodynamics, it is enough for our purpose to have considered $p$ and $\Delta$. We nevertheless mention for completeness that the entropy density, $s=\partial_T p$, has been independently computed on the lattice in Ref.~\cite{meyer} between 0.7$\, T_c$ and $T_c$. The present model actually overestimates the dimensionless ratio $s/T^3$ obtained in this last work, especially near $T_c$. We hope that more lattice data concerning the QCD EoS below $T_c$ will be available in the future in order to refine our model and perform a more quantitative comparison with the different lattice results. 

\section{Large $N_c$ limit}
Some remarks can be done about the large-$N_c$ limit of our model, that has not been considered up to now. By definition, $\kappa_{SB}\propto 1/(N^2_c-1)$. Above $T_c$, the dominant contribution to the equation of state should come from deconfined gluons, whose $(N^2_c-1)$ color degrees of freedom cancel the $\kappa_{SB}$ factor, leading to globally constant thermodynamical properties with respect to $N_c$ as observed in Ref.~\cite{panero}. Below $T_c$ however, all glueballs are in a color singlet and their mass scales as $N_c^0=O(1)$. The glueballs contribution to the EoS normalized according to $\kappa_{SB}$ should thus tend to zero at large $N_c$. But, since the gluon condensate scales as $(N^2_c-1)$, its thermodynamical contribution remains constant at large $N_c$. The trace anomaly and pressure at large $N_c$ are then expected to read $\Delta\approx \Delta_{G^2}$ and $p\approx p_{G^2}$ respectively in our model.

\begin{figure}[t]
\includegraphics*[width=8.0cm]{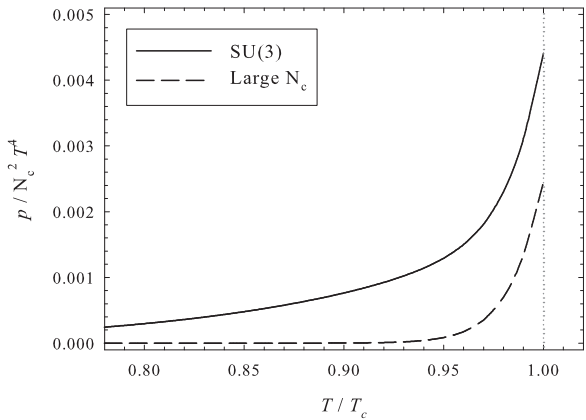}
\caption{Total pressure~(\ref{ptot}) normalized to $N_c^2T^4$ computed at $N_c=3$ (solid line) and at large $N_c$ (dashed line).}
\label{fig5}
\end{figure}

Following our scenario, the gluon condensate term is dominant in the trace anomaly but not in the pressure. Thus, we predict that the pressure below $T_c$ will be strongly suppressed at large $N_c$ as illustrated in Fig.~\ref{fig5}, while the trace anomaly will decrease only slightly. The results of Ref.~\cite{panero} suggest that the trace anomaly could be more and more peaked near $T_c$ at large $N_c$, leading to a phase transition which is more of first-order type, in qualitative agreement with what is expected from the present approach. Remark however that recently, a strong expansion of Yang-Mills theory at large $N_c$ has been proposed~\cite{lange}, in which the pressure is found to scale as $N_c^0$, with no $N_c^2$ term as in our model. It can be hoped that future studies of glueball and glueball condensate properties at finite-temperature and large $N_c$ will allow a more accurate validation of the ideas developed in this work. 

\section{Conclusion}\label{conclu}

A new way of understanding the pure glue QCD equation of state below $T_c$ has been proposed. The basic idea is that the pure glue hadronic matter consists in both a glueball gas and a gluon condensate part. It can be intuitively expected that the thermal width of glueballs tends to increase when approaching the deconfinement phase transition. Such a behavior is effectively taken into account in our formalism through a decrease of the glueball masses near $T_c$. Moreover, that effect has to be combined with the vanishing of the gluon condensate at the critical temperature. 

In order to illustrate qualitatively the proposed scenario, lattice data have been used as numerical inputs in the model: The glueball masses and the gluon condensate values have been taken from Refs.~\cite{suga1} and \cite{delia} respectively. Those data, when incorporated into computations, lead to an equation of state in good agreement with that computed in Ref.~\cite{panero}, thus providing an \textit{a posteriori} coherent interpretation of various independent existing results in finite-temperature QCD. 

Remark that, although the agreement reached by our model and the lattice data is excellent, the present approach aims at being a first step towards a more complete model. Its weakness is mostly the way of including the glueball decay width, \textit{i.e.} by redefining the glueball masses without modifying the form of the Bose-Einstein distribution, which can be questionable very close to $T_c$ where the decay widths become significant with respect to the glueball masses. The elaboration of a more realistic approach is left for future works.    

Finally, at large $N_c$, we predict a strong reduction and a weaker decrease of the pressure and trace anomaly below $T_c$ respectively. The present discussion suggests that new high-precision computations of glueballs and gluon condensate properties as well as of gluon plasma thermodynamics below $T_c$ might be worth in order to check the general scenario developed here and the suggested large-$N_c$ behavior in particular. 

\section*{Acknowledgments}
I thank the F.R.S.-FNRS for financial support. I also thank M. Panero for giving me his data, H. Suganuma for valuable comments about this work, as well as C. Semay, P. Castorina and E. Meggiolaro for useful discussions.

\end{document}